\documentclass{article}

\usepackage{microtype}
\usepackage{graphicx}
\usepackage{subcaption}
\usepackage{booktabs} 

\usepackage{hyperref}

\usepackage[accepted]{icml2026}

\usepackage{amsmath}
\usepackage{amssymb}
\usepackage{mathtools}
\usepackage{amsthm}
\usepackage{multirow}
\usepackage{tabularx}
\usepackage{array}

\usepackage[capitalize,noabbrev]{cleveref}

\theoremstyle{plain}

\theoremstyle{definition}

\theoremstyle{remark}

\usepackage[textsize=tiny]{todonotes}

\icmltitlerunning{Submission and Formatting Instructions for ICML 2026}

\begin{document}

\twocolumn[
    \icmltitle{Machine Learning Challenges in Intelligent Unmanned Aerial Vehicle Operations in Developing Economies}

  \begin{icmlauthorlist}
    \icmlauthor{Isuru Munasinghe}{uom}
    \icmlauthor{Nethmi Pathirana}{uom}
    \icmlauthor{Charitha Dombawala}{uom}
    \icmlauthor{Asanka Perera}{usq}
    \icmlauthor{Akila Pemasiri}{qut}
  \end{icmlauthorlist}

  \icmlaffiliation{uom}{Faculty of Engineering, University of Moratuwa, Katubedda, Sri Lanka}
  \icmlaffiliation{usq}{School of Engineering \& Digital Technologies, University of Southern Queensland, Brisbane, Australia}
  \icmlaffiliation{qut}{School of Electrical Engineering and Robotics, Queensland University of Technology, Brisbane, Australia}

  \icmlcorrespondingauthor{Isuru Munasinghe}{isuru.munasinghe1998@gmail.com}
  \icmlcorrespondingauthor{Akila Pemasiri}{a.thondilege@qut.edu.au}

  \icmlkeywords{Machine Learning, Unmanned Aerial Vehicle, Intelligent UAV Operations, Developing Economies}
    \vskip 0.1in
]

\printAffiliationsAndNotice{}

\begin{abstract}
Unmanned aerial vehicle (UAV) environments present significant challenges for machine learning (ML) due to limited platform resources, heterogeneous sensor data, dynamic mission conditions, and safety-critical requirements. This paper examines these constraints across the core functional areas of UAV intelligence, including navigation, perception, communication-aware operation, and resilience specifically in the context of developing economies. In such settings, these challenges are often amplified by constraints such as cost sensitivity, limited infrastructure, intermittent connectivity, regulatory uncertainty, and harsh or variable operating environments. The discussion highlights the gap between ML performance in controlled experimental backgrounds and dependable deployment in real-world UAV missions within developing economies context.
\end{abstract}

\vspace{-1cm}
\section{Introduction}
\vspace{-1mm}

\begin{figure*}[t]
    \centering
    \includegraphics[width=0.95\textwidth]{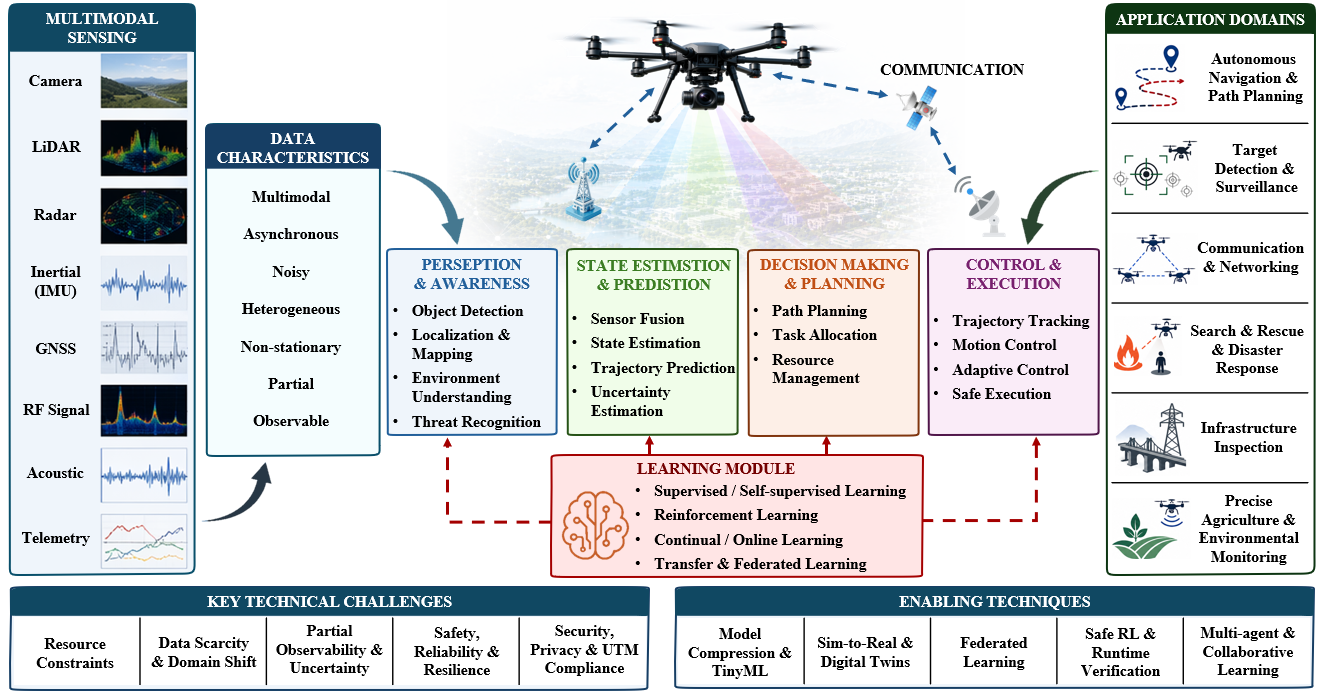}
    \caption{Overview of ML-based UAV operations, highlighting key functional modules, application domains, and technical challenges.}
    \label{overall}
\end{figure*}

Recent advances in ML have accelerated intelligent UAV systems for environmental monitoring, infrastructure inspection, search and rescue, and communication support. These systems are especially valuable in developing countries for critical operations in resource-limited and hard-to-access areas. However, reliable deployment remains challenging due to limited onboard computation, energy constraints, dynamic flight conditions, and tightly connected sensing, communication, perception, and control requirements~\cite{shah2024applications, komatineni2024comprehensive, daoud2025investigating}. These challenges are further intensified in low-income economies, where affordability constraints often necessitate the use of UAV platforms with even smaller size, weight, and power (SWaP) budgets, significantly restricting available sensing, processing, and communication resources and placing additional pressure on ML model efficiency and robustness~\cite{singh2025review}.

From a technical perspective, UAV-oriented ML problems extend beyond standard prediction tasks. They involve state estimation, trajectory optimization, perception, network-aware decision-making, and fault-resilient operation. Noisy and asynchronous sensor data, domain shift across environments, and uncertainty in state estimation can reduce model generalization and deployment stability. In safety-critical UAV operations, ML models must satisfy strict requirements for latency, robustness, and interpretability, since failures can directly affect mission success and operational safety.

The main contribution of this paper is the organization of UAV-oriented ML problems into key functional domains, including navigation, perception, communication-aware intelligence, and operational resilience, as illustrated in Fig.~\ref{overall}. The paper discusses major deployment limitations in UAV environments, with an emphasis on resource-aware inference, robust generalization, multimodal perception, safety assurance, feasibility of low-resource deployment, and unmanned aircraft system traffic management (UTM)-compliant operation~\cite{hamissi2023survey}.

\vspace{-3mm}
\section{ML-Based Framework for Intelligent UAV Operations}
\vspace{-1mm}

This section presents a technical assessment of the ML-based framework for intelligent UAV operations. It examines practical deployment conditions in developing countries and low-economic regions, along with the principal ML problem domains, fundamental methodological approaches, and major technical challenges that constrain reliable and scalable real-world deployment.

\vspace{-3mm}
\subsection{Operational Characteristics and ML Problem Domains in UAV Systems}
\vspace{-1mm}

The formulation of ML problems for intelligent UAV operations is mainly influenced by platform architecture, sensing configuration, mission requirements, communication availability, and airspace integration. As illustrated in Fig.~\ref{overall}, intelligent UAV operation depends on the coordinated use of perception, mobility, communication, and resilience-related functions. In developing economies and resource-constrained regions, these requirements are further affected by limited infrastructure, restricted access to advanced hardware, scarce local datasets, and reduced maintenance capacity. Therefore, UAV-oriented ML must be evaluated through both algorithmic performance and practical deployment feasibility. The principal challenges and research directions associated with UAV-oriented ML are summarized in Table~\ref{uav_ml_challenges}.

\begin{table*}[!t]
\centering
\caption{Principal technical challenges and research directions in UAV-oriented ML.}
\label{uav_ml_challenges}
\renewcommand{\arraystretch}{1.12}
\setlength{\tabcolsep}{4pt}
\footnotesize
\begin{tabularx}{\textwidth}{
>{\raggedright\arraybackslash}p{2.6cm}
>{\raggedright\arraybackslash}p{7.4cm}
>{\raggedright\arraybackslash}p{6.3cm}}
\hline
\textbf{Challenge domain} &
\textbf{ML difficulty and operational impact} &
\textbf{Technical directions} \\
\hline

\textbf{Resource-constrained onboard inference}
& Limited battery, payload, memory, thermal budget, and embedded processing restrict real-time convolutional neural network (CNN), recurrent neural network (RNN), transformer, and deep reinforcement learning (DRL) deployment.
& Lightweight models, pruning, quantization, knowledge distillation, Tiny machine learning (TinyML), edge-assisted inference, and computation offloading~\cite{xia2023survey} \\
\hline

\multirow{2}{=}{\textbf{Data scarcity and domain shift}}
& Limited labeled data across UAV platforms, sensors, weather, and mission scenarios weakens field generalization.
& Transfer learning, domain adaptation, synthetic data, self-supervised learning \\
\cline{2-3}
& Non-stationary distributions from altitude, terrain, illumination, wind, occlusion, and sensor degradation reduce prediction reliability.
& Continual learning, online calibration, uncertainty estimation, adaptive normalization \\
\hline

\multirow{2}{=}{\textbf{Real-time operation under partial observability}}
& UAV states are inferred from noisy asynchronous inertial, visual, radio frequency (RF), radar, acoustic, and telemetry streams.
& Sensor fusion, Bayesian filtering, long short-term memory (LSTM), gated recurrent unit (GRU), partially observable Markov decision process (POMDP)-based learning \\
\cline{2-3}
& Moving obstacles, changing links, and incomplete maps increase replanning complexity and reduce static policy reliability.
& DRL, learned model predictive control (MPC), graph-based planning, risk-aware trajectory optimization \\
\hline

\multirow{2}{=}{\textbf{Robust multimodal perception}}
& RF, visual, acoustic, and radar modalities fail under different interference, occlusion, noise, and low radar cross-section (RCS) conditions~\cite{zhu2024intelligent}.
& Multimodal fusion, attention-based sensor weighting, confidence-aware prediction \\
\cline{2-3}
& Synchronization error, calibration drift, and heterogeneous sampling reduce consistency in detection and localization.
& Time-aligned fusion, sensor reliability estimation, multimodal transformers, late-fusion decisions \\
\hline

\textbf{Safety-critical reliability and recovery}
& Hardware faults, sensor failures, communication loss, actuator degradation, and cyberphysical attacks can destabilize UAV operation.
& Fault prediction, anomaly detection, safe recovery learning, resilient control, runtime assurance~\cite{adaika2025fault} \\
\hline

\multirow{2}{=}{\textbf{Security, privacy, and UTM-compliant operation}}
& Beyond-visual-line-of-sight (BVLOS) and UTM-assisted operations require secure identification, conformance monitoring, trajectory sharing, and traceability.
& Privacy-preserving learning, federated learning, secure remote identification (RID) analytics, UTM-aware decision making \\
\cline{2-3}
& Spoofing, jamming, eavesdropping, and adversarial perturbations threaten navigation integrity and model trust.
& Adversarially robust ML, secure fusion, intrusion detection, authentication-aware learning \\
\hline
\end{tabularx}
\vspace{-2mm}
\end{table*}

\vspace{-3mm}
\subsubsection{Platform, Payload, and Mission Constraints}
\vspace{-1mm}

UAV platforms operate under strict aerodynamic, energetic, computational, and communication constraints. Flight stability, mission execution, sensing, and onboard inference must be maintained within limited SWaP and cost budgets. As a result, the suitability of an ML architecture depends on the vehicle type, onboard payload capacity, sensing stack, navigation requirements, and the economic and infrastructural conditions of the deployment region~\cite{mcenroe2022survey}.

The payload subsystem introduces a major design constraint in intelligent UAV operations. Cameras, light detection and ranging (LiDAR) units, radar front ends, inertial measurement units, global navigation receivers, communication transceivers, and embedded compute modules must operate within limited payload and power budgets. Therefore, onboard ML capability cannot be selected based on predictive accuracy alone. It must also account for sensing quality, computational throughput, inference latency, thermal behavior, communication overhead, and mission duration.

These constraints are more prominent in developing economies and resource-constrained regions, where access to high-performance sensors, embedded processors, maintenance facilities, and replacement components is often limited. This increases the importance of mission-aware and resource-efficient ML models that can operate reliably within practical airframe, payload, and deployment limitations.

\vspace{-3mm}
\subsubsection{Navigation, Perception, and Data Characteristics}
\vspace{-1mm}

UAV navigation depends on inertial, satellite-based, vision-based, and communication-assisted positioning methods, each affected by drift, signal blockage, spoofing, occlusion, and connectivity variation. In practical operations, these methods are commonly integrated through multisensor fusion for reliable state estimation, especially in BVLOS missions where direct human supervision is unavailable~\cite{theile2020uav}. The resulting data streams are multimodal, asynchronous, noisy, and non-stationary, including inertial measurements, visual frames, RF signatures, radar returns, telemetry, and mission-state variables.

Autonomous navigation and trajectory optimization represent central ML problem domains in UAV systems. Path planning, collision avoidance, waypoint tracking, and mission execution are commonly formulated as optimization or sequential decision-making tasks. Classical graph-based and optimization methods have been widely applied for feasible trajectory generation, while recent studies increasingly use reinforcement learning (RL), DRL, and vision-driven policies for navigation in dynamic and partially observable environments~\cite{ekechi2025survey, liu2024deep}.

Perception-oriented ML contributes to localization, mapping, object identification, obstacle awareness, target detection, and environmental interpretation~\cite{tang2023survey, leng2024recent}. Representative approaches include CNNs, RNNs, support vector machines (SVMs), You Only Look Once (YOLO) variants, Faster region-based convolutional neural network (Faster R-CNN), Cascade R-CNN, and radar spectrogram-based architectures~\cite{zhang2024lightweight}. In developing economies, degraded global navigation satellite system (GNSS) signals, limited high-resolution maps, unstructured terrain, low-cost sensing platforms, restricted onboard computation, and scarce annotated local datasets reduce localization accuracy, model generalization, and validation quality, as reflected in Table~\ref{uav_ml_challenges}.

\vspace{-3mm}
\subsubsection{Communication-Aware Intelligence and Operational Resilience}
\vspace{-1mm}

Communication-aware and network-assisted UAV intelligence considers UAVs as aerial communication nodes, relays, base stations, or user equipment within larger wireless systems. ML methods are used for channel modeling, interference mitigation, user association, placement optimization, spectrum allocation, power management, and trajectory-aware communication control~\cite{sun2024advancing}. Representative approaches include artificial neural networks (ANNs), k-nearest neighbors (KNN), RL, Gaussian mixture models (GMMs), liquid state machines (LSMs), and multi-agent Q-learning~\cite{amodu2025comprehensive}. This domain is important because communication reliability directly affects mission planning, data transmission, and real-time decision-making.

Failure diagnosis and operational resilience address the detection and mitigation of hardware faults, sensor failures, communication loss, actuator faults, cyberphysical disruption, and UTM-level anomalies~\cite{fang2025fault, roshanski2024real}. In developing economies, limited connectivity, weak maintenance infrastructure, shortage of skilled personnel, delayed spare-part access, and harsh environmental conditions can reduce system resilience and increase operational risk~\cite{wan2024advancements}. Therefore, sustained UAV operation depends on ML reliability, maintainability, deployment feasibility, and the ability to adapt to degraded operating conditions.

\vspace{-3mm}
\subsubsection{Resource-Aware Challenges in UAV-Oriented ML Deployment}
\vspace{-1mm}

The practical realization of ML in UAV systems introduces several interrelated challenges that extend beyond conventional predictive modeling, as summarized in Table~\ref{uav_ml_challenges}. These limitations are more prominent in developing countries and low-economic regions, where access to advanced onboard processors, reliable communication infrastructure, maintenance support, and locally representative datasets may be restricted by economic and infrastructural constraints. In addition, UAV operation often occurs under partial observability, where noisy and asynchronous multimodal sensor measurements introduce uncertainty that can affect subsequent decision-making.

Addressing these challenges requires adaptive and system-aware learning approaches that balance predictive performance with practical deployment constraints. Resource-aware optimization can reduce computational overhead by adjusting model complexity and inference frequency according to mission requirements, available energy, and onboard processing capacity. This is especially important in developing countries and resource-constrained regions, where UAV systems often operate with limited hardware and limited access to representative datasets. Generalization and reliability can be improved through simulation-based pretraining, self-supervised learning, continual adaptation to local conditions, uncertainty-aware multimodal fusion, safe learning methods, runtime verification, and fallback control policies under disturbances, communication loss, and resource limitations.

\vspace{-3mm}
\subsection{Non-Technical Challenges in UAV-Oriented ML in Developing Economies}
\vspace{-1mm}

Although ML techniques for UAV systems have advanced considerably, their practical deployment in developing economies is often constrained by factors beyond algorithmic performance. Economic limitations, infrastructure readiness, regulatory maturity, public acceptance, and operational sustainability can significantly influence the adoption and scalability of UAV-based solutions~\cite{upadrasta2025public}. These factors are important in resource-constrained environments, where technical performance alone may not ensure long-term deployment feasibility.

\textbf{Economic Constraints:}
Cost remains a major barrier to UAV deployment in developing economies. Limited funding can restrict the acquisition of high-quality UAV platforms, advanced sensors, embedded computing modules, and supporting software systems. Long-term operation also requires continuous investment in maintenance, staff training, system upgrades, and field deployment logistics. Therefore, UAV-based ML solutions must be evaluated not only in terms of predictive performance but also in terms of affordability, operational cost, and scalability.

\textbf{Infrastructure Limitations:}
Infrastructure limitations can directly affect the reliability of UAV operations. In many developing regions, unstable internet connectivity, limited access to electricity, weak communication networks, and insufficient ground support facilities can restrict real-time data transmission, remote monitoring, cloud-based processing, and coordinated UAV operation. These limitations are especially critical in rural and remote areas, where UAV applications such as disaster assessment, agricultural monitoring, and infrastructure inspection are often most valuable~\cite{epifani2024survey}.

\textbf{Regulatory and Policy Challenges:}
The regulatory environment for UAV operations is still developing in many emerging economies~\cite{kemarau2024global}. Unclear policies related to airspace access, licensing, flight permissions, safety requirements, data governance, and BVLOS operation can create uncertainty for organizations seeking to deploy UAV systems. The absence of mature UTM frameworks can further limit large-scale and coordinated UAV deployment.

\textbf{Social Acceptance and Public Perception:}
Public acceptance is an important factor in the successful adoption of UAV technologies. Concerns related to privacy, surveillance, safety, noise, and misuse of collected data may reduce community trust in UAV-based systems. Limited public awareness of the benefits and intended uses of UAV technology can further affect adoption, particularly in populated or sensitive areas. Therefore, transparent communication, responsible data handling, and community engagement are important for improving acceptance.

\textbf{Maintenance and Operational Sustainability:}
The long-term reliability of UAV systems depends on the availability of maintenance services, replacement components, skilled operators, and technical support. In developing economies, limited access to these resources can increase downtime and reduce operational sustainability. Environmental conditions such as dust, humidity, high temperature, and heavy rainfall can further increase maintenance requirements and affect sensor and platform reliability.

Therefore, successful UAV deployment in developing economies depends on ML model capability, economic feasibility, infrastructure readiness, regulatory support, public acceptance, and sustainable maintenance practices.

\vspace{-3mm}
\section{Conclusion}
\vspace{-1mm}

This paper presented a technical analysis of ML-based approaches for intelligent UAV operations by examining the relationship between platform constraints, sensing characteristics, and operational requirements. The discussion covered major ML problem domains, including navigation, perception, communication-aware intelligence, and operational resilience, identifying key challenges such as resource-constrained inference, domain shift, partial observability, and safety-critical reliability. The analysis shows that effective UAV intelligence requires system-aware learning approaches that address efficiency, reliability, and adaptability while also considering the cost, connectivity, maintenance, and data-availability constraints common in developing countries and low-resource economies.

\bibliographystyle{icml2026}
\bibliography{references}

\end{document}